\documentclass[aps,preprint,amsmath,amssymb]{revtex4}
\usepackage{epsfig}
\usepackage{color}
\usepackage{bm}
\usepackage{graphicx}
\usepackage{subfigure}

\begin{document}
\draft
\title{
{\bf Canonical Charmonium Interpretation for Y(4360) and Y(4660)}}

\author{Gui-Jun Ding$^{a}$}
\author{Jie-Jie Zhu$^{a}$}
\author{ Mu-Lin Yan$^{a,b}$}

\affiliation{\centerline{$^a$Department of Modern
Physics,}\centerline{University of Science and Technology of
China,Hefei, Anhui 230026, China}\centerline{$^b$ Interdisciplinary
Center for Theoretical Study,} \centerline{University of Science and
Technology of China,Hefei, Anhui 230026, China}}

\begin{abstract}
In this work, we consider the canonical charmonium assignments for
Y(4360) and Y(4660). Y(4660) is good candidate of $\rm 5\,^3S_1$
$c\bar{c}$ state, the possibility of Y(4360) as a $\rm 3\,^3D_1$
$c\bar{c}$ state is studied, and the charmonium hybrid
interpretation of Y(4360) can not be excluded completely. We
evaluate the $e^{+}e^{-}$ leptonic widths, E1 transitions, M1
transitions and the open flavor strong decays of Y(4360) and
Y(4660). Experimental tests for the charmonium assignments are
suggested.

\vskip 0.5cm


\end{abstract}
\maketitle
\section{introduction}
In the last five years we have witnessed a revival of interest in
charm spectroscopy, the B-factories( Babar and Belle ) and other
machines have reported a large number of new states with hidden
charm: $h_c$(1P)\cite{hc}, $\eta_c$(2S)\cite{etac},
$\rm{X}$(3872)\cite{X3872}, $\rm{X}$(3940)\cite{X3940},
$\rm{Y}$(3940)\cite{Y3940}, $\rm{Z}$(3930)\cite{Z3930} and
$\rm{Y}$(4260)\cite{Y4260}. Some of them can be understood as
$c\bar{c}$ states, while a conventional assignments for some are
elusive( for a recent review, see, e.g.\cite{review} ). These
discoveries are enriching and also challenging our knowledge for the
hadron spectroscopy, and the underlining theory for strong
interactions.

Recently the Belle collaboration has observed two charmonium-like
states Y(4360) and Y(4660) in
$e^{+}e^{-}\rightarrow\pi^{+}\pi^{-}\psi(2S)$ via initial state
radiation\cite{Y4360}. The mass of Y(4360) is
$4361\pm9\pm9\rm{MeV}/c^2$ with a width of
$74\pm15\pm10\rm{MeV}/c^2$ and the statistical significance is of
more than $8\sigma$. The mass of Y(4660) is
$4664\pm11\pm5\rm{MeV}/c^2$ with a width of
$48\pm15\pm3\rm{MeV}/c^2$ and statistical significance $5.8\sigma$.
Both these two structures are known to be produced in initial state
radiation from $e^{+}e^{-}$ annihilation and hence to have ${\rm
J^{PC}=1^{--}}$. They were seen in the  decays with products of
$\pi^{+}\pi^{-}\psi(2S)$. It has been determined by the Belle
collaboration that
\begin{eqnarray} \nonumber \Gamma(Y(4360)\rightarrow
e^{+}e^{-}){\cal
B}(Y(4360)\rightarrow\pi^{+}\pi^{-}\psi(2S))&=&10.4\pm1.7\pm1.5(
11.8\pm1.8\pm1.4)\rm{eV/c^2}\\
\label{1}\Gamma(Y(4660)\rightarrow e^{+}e^{-}){\cal
B}(Y(4660)\rightarrow\pi^{+}\pi^{-}\psi(2S))&=&3.0\pm0.9\pm0.3(
7.6\pm1.8\pm0.8)\rm{eV/c^2}
\end{eqnarray}
where the numbers in the bracket are the solution II fit performed
by Belle. In order to understand the nature of Y(4360) and Y(4660),
it is worth to note that the Babar collaboration has observed a
broad structure Y(4325) in the process
$e^{+}e^{-}\rightarrow\gamma_{ISR}\pi^{+}\pi^{-}\psi(2S)$ at
$4324\pm24\rm{MeV/c^2}$ with a width
$172\pm33\rm{MeV/c^2}$\cite{Y4325}. The mass of Y(4360) is close to
that of Y(4325), the main difference between Y(4325) and Y(4360) is
their widths, and it seems very difficult to observe these two
structures simultaneously because of the large width of Y(4325). If
both Y(4325) and Y(4360) are not experimental artifacts, they could
be the same structure and the width difference is due to the
experimental error, or they are two different resonances. If one
assumed that they were two different  structures, it would be very
difficult to assign both of them as conventional charmoniums
simultaneously. So, at least one of them should be exotic state. It
possibly may be produced by ${\rm D_1\overline{D}}$( or {\rm $\rm
D_{s1}\overline{D}_s$} ) rescattering effect or some other
mechanism, therefore it would indicate the necessity of refinements
in the naive "quenched" ${\rm q\overline{q}}$ quark models or the
inclusion of additional dynamical effects. In this paper, we assume
Y(4325) and Y(4360) are exactly the same resonance for simplicity.
Possible non-charmonium assignment of Y(4360) and the relation
between Y(4325) and Y(4360) will be considered in detail in future
work. Finally we would like to mention that the Belle collaboration
claims the broad Y(4325) is comprised of two narrower peaks Y(4360)
and Y(4660)\cite{bellelp}.

In order to understand the structure of Y(4360) and Y(4660), i.e.
whether they are conventional charmonium states or other exotic
structures, it is very necessary to first consider the canonical
charmonium assignments and the characteristic signals. With ${\rm
J^{PC}=1^{--}}$, a conventional $c\bar{c}$ state is either a S-wave
state or a D-wave state. There are already reasonably well
established $c\bar{c}$ candidates for 1S, 2S, 1D, 3S, 2D and
4S\cite{pdg}, therefore new $1^{--}$ charmonium states can only
belong to 3D, 4D, 5S or 6S, and a natural assignment for Y(4360)
will be a ${\rm 3\,^3D_1}$ $c\bar{c}$ state, and Y(4660) as a ${\rm
5\,^3S_1}$ charmonium. However, this assignment has the problem that
the mass of the Y(4360) is somewhat lower than the nonrelativistic
potential model prediction for the ${\rm 3\,^3D_1}$ $c\bar{c}$
state, which will be shown in the next section.

In this work, we study the properties of Y(4360) and Y(4660) under
the hypothesis of Y(4360) as a ${\rm 3\,^3\rm{D}_1}$ $c\bar{c}$
state and Y(4660) as ${\rm 5\,^3\rm{S}_1}$ $c\bar{c}$ state. We
first briefly review the nonrelativistic potential model and give
its prediction for the masses of ${3\,^3\rm{D}_1}$,
${4\,^3\rm{D}_1}$, ${5\,^3\rm{S}_1}$ and ${6\,^3\rm{S}_1}$
$c\bar{c}$ states. The $e^{+}e^{-}$ leptonic widths, E1 transitions,
M1 transitions and open charm strong decays of both Y(4360) and
Y(4660) are studied in section III and IV respectively. From these
results, we suggest adequate measurements which can verify the
canonical charmonium assignments and distinguish the $c\bar{c}$
structure from other non-$c\bar{c}$ possibilities. Finally we
present our summary and some discussions. Possible charmonium hybrid
assignment of Y(4360) and its crucial decay modes are suggested.

\section{Review on nonrelativistic potential model and the charmonium
assignments for Y(4360) and Y(4660) }

The quark potential models have successfully described the
charmonium spectrum, which generally assumes shorted-ranged color
coulomb interaction and long-ranged linear scalar confining
interaction plus spin dependent part coming from one gluon exchange
and the confining interaction.The potental model is closely related
to QCD, which can be derived from the QCD effective field
theory\cite{Brambilla:2004jw,Brambilla:1999ja}. Here we shall use
the simple nonrelativistic potential model proposed by T.Barnes and
S.Godfrey and E.S.Swanson\cite{bgs}, the zeroth-order Hamiltonian
is,
\begin{equation}
\label{2}\rm{H}_0=\frac{\mathbf{p}^{\;2}}{
\;m_c}-\frac{4}{3}\frac{\alpha_s}{r}+br+\frac{32\pi\alpha_s}{9m^2_c}
\;\tilde{\delta}_{\sigma}(r)\,\mathbf{S}_c\cdot\mathbf{S}_{\bar{c}}
\end{equation}
where
$\tilde{\delta}_{\sigma}(r)=(\sigma/\sqrt{\pi})^3\;e^{-\sigma^2r^2}$,
which is a gaussian-smeared hyperfine interaction. Solution of the
Schr$\ddot{\rm{o}}$dinger equation with the above $\rm{\rm{H}}_0$
gives our zeroth order charmonium wavefunctions. The splitting
within the multiplets is then determined by taking the matrix
element of the spin-dependent Hamiltonian $\rm{H}_{sd}$ between
these zeroth-order wavefunctions. The spin-dependent Hamiltonian is
taken from the one-gluon-exchange Breit-Fermi Hamiltonian (which
gives spin-orbit and tensor terms) and an inverted spin-orbit term,
which follows from the assumption of a Lorentz scalar confining
interaction. The $\rm{H}_{sd}$ is as follows,
\begin{equation}
\label{3}\rm{H}_{sd}=\frac{1}{m^2_c}\;[(\frac{2\alpha_s}{r^3}-\frac{b}{2r})\;\mathbf{L}\cdot\mathbf{S}+
\frac{4\alpha_s}{r^3}\mathbf{T}]
\end{equation}

This simple potential model consists of four parameters: the strong
coupling constant $\alpha_s$ which is taken to be a constant for
simplicity, the string tension $b$, the charm quark mass $m_c$, and
the hyperfine interaction smearing parameter $\sigma$. Fitting the
masses of the 11 reasonably well established experimental charmonium
states, the values of these four parameters are already fixed as
follows: $\alpha_s=0.5461$, $b=0.1425\rm{GeV}^2$,
$m_c=1.4794\rm{GeV}$ and $\sigma=1.0946\rm{GeV}$\cite{bgs}. Solving
the Schr$\ddot{\rm{o}}$dinger equation with the zeroth-order
Hamiltonian $\rm{H}_{0}$ numerically by the Mathematica
program\cite{solves} and treating the spin-dependent terms
$\rm{H}_{sd}$ as mass shifts by the leading order perturbation, we
obtain the masses and wavefuntions of the canonical $c\bar{c}$
states. The masses of $3\,^3\rm{D}_1$, $4\,^3\rm{D}_1$,
$5\,^3\rm{S}_1$ and $6\,^3\rm{S}_1$ are predicted as,
\begin{eqnarray}
\nonumber&&\rm{M}(3\,^3\rm{D}_1)=4455\rm{MeV},~~~\rm{M}(4\,^3\rm{D}_1)=4740\rm{MeV},\\
\label{4}&&\rm{M}(5\,^3\rm{S}_1)=4704\rm{MeV},~~~\rm{M}(6\,^3\rm{S}_1)=4977\rm{MeV}
\end{eqnarray}

Comparing with the masses of Y(4360) and Y(4660), it is natural to
assign Y(4360) as a $3\,^3\rm{D}_1$ and Y(4660) as a $5\,^3\rm{S}_1$
canonical charmonium states. Although the mass of Y(4360) is
somewhat smaller than the theoretical prediction, however, we notice
that the mass predictions of various potential model for the high
charmonium may differ by 10-100\rm{MeV}\cite{review}, therefore
Y(4360) as a $3\,^3\rm{D}_1$ $c\bar{c}$ state is not irrational. In
this work we assume that the discrepancy in the spectrum is due to
the theoretical uncertainties or other effects such as the coupled
channel effects. It is interesting to refit the parameters
$\alpha_s$, $b$, $m_c$ and $\sigma$ including both Y(4360) and
Y(4660) or only Y(4660).

\section{electronmagnetic transitions of Y(4360) and Y(4660)}

\subsection{ The $e^{+}e^{-}$ leptonic decay of Y(4360) and Y(4660)}

The decay of quarkonium state into a lepton pair proceeds via a
single virtual photon, as long as the mass of the initial quarkonim
is sufficiently small that the contribution of a virtual $Z$ can be
ignored. The leptonic partial decay widths probe the compactness of
the quarkonium system, and they provide useful information about the
wavefunctions of the $1^{--}$ quarkonium states. The leptonic width
of $n\,^3\rm{S}_1$ charmonium is given by\cite{rw,quigg},
\begin{equation}
\label{5}\Gamma(n^3{\rm{S}}_1\rightarrow
e^+e^-)=\frac{4\alpha^2e^2_c|\psi_n(0)|^2}{\rm{M}^2_n}(1-\frac{16\alpha_s}{3\pi}+...)
\end{equation}
where $e_c=2/3$ is the charm quark electric charge, $\rm{M}_n$ is
the mass of the $n\,^3\rm{S}_1$ state, and the second term is the
QCD correction. $\psi_n(0)$ is the $n\,^3\rm{S}_1$ wavefunction at
the origin, and the radial wavefunction is normalized according to
$\int_0^{\infty}dr\;r^2|\psi_n(r)|^2=1$. At the leading order, the
width of D-wave $c\bar{c}$ states to $e^{+}e^{-}$ is proportional to
$|\psi_n^{''}(0)|^2$,
\begin{equation}
\label{6}\Gamma(n^3{\rm{D}}_1\rightarrow
e^+e^-)=\frac{25\alpha^2e^2_c}{2m^4_c\rm{M}^2_n}|\psi^{''}_n(0)|^2
\end{equation}
which is generally smaller than the corresponding widths of the
$n\,^3{\rm S}_1$ states. Using the nonrelativistic quark model
wavefunctions calculated in the previous section, we evaluate these
leptonic decay widths of Y(4360) and Y(4660) at both the
experimental values and the theoretical predictions of
nonrelativistic potential model. The width predictions are given in
Table \ref{leptonic}, where we choose
$\alpha_s\approx0.23$\cite{pdg}.

\begin{table}[hptb]
\caption{\label{leptonic} The $e^{+}e^{-}$ partial widths of Y(4360)
and Y(4660).}
\begin{ruledtabular}
\begin{tabular}{lcc}
Initial state & Mass (GeV) & $\Gamma_{e^+e^-}$ (keV)\\\hline
Y(4360)($3^3\rm{D}_1)$     & 4.361 & 0.87\\
                           & 4.455 & 0.83\\
\hline
Y(4660)($5^3\rm{S}_1$)     & 4.664 & 1.34\\
                           & 4.704 & 1.32
\end{tabular}
\end{ruledtabular}
\end{table}

Using Eq.(\ref{1}), we can estimate that
\begin{eqnarray}
\nonumber {\cal
B}(Y(4360)\rightarrow\pi^{+}\pi^{-}\psi(2S))&\sim&1.20\times10^{-2}
(\;\rm{or}\;1.36\times10^{-2} )\\
\label{7}{\cal
B}(Y(4660)\rightarrow\pi^{+}\pi^{-}\psi(2S))&\sim&2.24\times10^{-3}
(\;\rm{or}\;5.67\times10^{-3})
\end{eqnarray}

The numbers in the bracket are the results corresponding to the
solution II fit by the Belle collaboration. Generally we expect the
branch fraction for
$c\bar{c}(3\,^3\rm{D}_1)\rightarrow\pi^{+}\pi^{-}\psi(2S)$ should be
of the order $10^{-3}$, e.g., ${\cal
B}(\psi(4160)\rightarrow\pi^{+}\pi^{-}\psi(2S))<4\times10^{-3}$\cite{pdg},
where $\psi(4160)$ is a good candidate of $2\,^3\rm{D}_1$ $c\bar{c}$
state. Therefore ${\cal
B}(Y(4360)\rightarrow\pi^{+}\pi^{-}\psi(2S))$ seems a little larger,
which may be because of the QCD radiative corrections to
$\Gamma({\rm Y}(4360)\rightarrow e^{+}e^{-})$, and non-valence
components may also contribute, which deserves investigating
further. ${\cal B}(Y(4660)\rightarrow\pi^{+}\pi^{-}\psi(2S))\sim
10^{-3}$, which indicates that Y(4660) may be a good candidate of
$5\,^3\rm{S}_1$ $c\bar{c}$ state.

\subsection{Radiative transitions of Y(4360) and Y(4660)}

Radiative decay of higher-mass charmonium states is an important way
to produce lower charmonium states, and it plays significant role in
charmonium physics. By means of the radiative transitions one can
probe the internal charge structure of hadrons, hence it is useful
for determining the quantum numbers and hadronic structures of heavy
quark mesons. The radiative transition amplitude is determined by
the matrix element of the EM current between the initial quarkonium
state $i$ and the final state $f$, i.e., $\langle
f|j^{\mu}_{em}|i\rangle$. Expanding in powers of photon momentum
generates the electric and magnetic multipole moments, the leading
order transition amplitudes are electric dipole (E1) transion or
magnetic dipole (M1) transition. They are quite straightforward to
be evaluated in the potential model.

\subsubsection{E1 transitions}

The partial width for E1 transitions between states ${\rm
n}\,^{2{\rm S}+1}{\rm L}_{\rm J}$ and ${\rm n}'\,^{2{\rm S}'+1}{\rm
L}'_{{\rm J}'}$ $c\bar{c}$ state in the nonrelativistic quark model
is given by\cite{Eichten:1974af,
Eichten:1975ag,Kwong:1988ae,Brambilla:2004wf},
\begin{equation}
\label{8}\Gamma_{{\rm E1}}({\rm n}^{2{\rm S}+1}{\rm L}_{\rm
J}\rightarrow {\rm n}'\,^{2{\rm S}'+1}{\rm L}'_{{\rm
J}'}+\gamma)=\frac{4\alpha e^2_c{\rm E}^3_{\gamma}}{3}(2{\rm
J}'+1){\cal S}_{fi}\,\delta_{\rm S,S^{'}}|\langle{\rm n}'\,^{2{\rm
S'}+1}{\rm L}'_{{\rm J}'}|r|{\rm n}\,^{2{\rm S}+1}{\rm L}_{\rm
J}\rangle|^2\frac{{\rm E}_f}{{\rm M}_i}
\end{equation}
where ${\rm E}_{\gamma}$ is the photon energy, ${\rm E}_f$ is the
energy of final state ${\rm n}'\,^{2{\rm S}'+1}{\rm L}'_{{\rm J}'}$,
and ${\rm M}_i$ is the mass of the initial state ${\rm n}^{2{\rm
S}+1}{\rm L}_{\rm J}$. We have included the relativistic phase
factor $\frac{{\rm E}_f}{{\rm M}_i}$, and the statistical factor
${\cal S}_{fi}$ is
\begin{equation}
\label{9}{\cal S}_{fi}={\rm max}({\rm L}, {\rm L'})\cdot\left\{
\begin{array}{ccc}
{\rm L}'& {\rm J}' &{\rm S}\\
{\rm J} & {\rm L}  &1
\end{array}\right\}^2
\end{equation}

The matrix element $\langle{\rm n}'\,^{2{\rm S}'+1}{\rm L}'_{{\rm
J}'}|r|{\rm n}\,^{2{\rm S}+1}{\rm L}_{\rm J}\rangle$ can be
straightforwardly evaluated using the nonrelativistic
Schr$\ddot{\rm{o}}$dinger wavefunctions of the model described in
the previous section, and the resulting E1 transition widths of
Y(4360) and Y(4660) together with the photon energies are given in
Table \ref{E1Y4361}-- Table \ref{E1Y4704}, where the E1 transition
widths predictions for initial state assuming both the experiment
observed masses and the nonrelativistic potential model predictions
are given. The masses of the involved final state charmoniums are
taken from the Particle Data Group\cite{pdg} if the state is
included in the meson summary table. If it is not, then the masses
predicted in the nonrelativistic potential model described in the
previous section are used. The exceptions are $h_c(^1{\rm P}_1)$ and
$\eta_c(3^1{\rm S}_0)$, we assume ${\rm M}(\eta_c(3\,^1{\rm
S}_0))=4.011{\rm GeV}$(the mass of the known $\psi(4040)$ minus the
theoretical 3S splitting) and ${\rm M}(h_c(^1{\rm P}_1))=3.525{\rm
GeV}$, which is the spin-averaged mass of the $^3{\rm P}_{\rm J}$
$\chi_{\rm J}$ states.

From Table \ref{E1Y4361} and Table \ref{E1Y4455}, we can see that
Y(4360) should have very small E1 radiative widths to the triplet
member of the 1P multiplet and $1\,^3{\rm F}_2$ state, if Y(4360) is
a pure $3\,^3{\rm D}_1$ state. The radiative widths to the unknow 3P
and 2P triplet states $\chi_0(3\,^3{\rm P}_0)$, $\chi_1(3\,^3{\rm
P}_1)$, $\chi_0(2\,^3{\rm P}_0)$ and $\chi_1(2\,^3{\rm P}_1)$ are
theoretically large, so the radiative decays of Y(4360) can be used
to produce these states. Since the structures of both X(3940) and
Y(3940) are still unclear, they possibly belong to the $2\,^3{\rm
P}_{\rm J}$ multiplet\cite{X3940,Y3940}. Consequently the E1
transitions of Y(4360) into $\chi_0(2^3{\rm P}_0)$, $\chi_1(2^3{\rm
P}_1)$ are especially of interests, which maybe helpful to
clarifying the issue of X(3940) and Y(3940).

Next we consider the E1 transition of Y(4660) as a $5\,^3{\rm S}_1$
$c\bar{c}$ state. As is shown evidently in Table \ref{E1Y4664} and
Table \ref{E1Y4704}, the strong suppression of Y(4660) E1 decays to
${\rm n}^3{\rm P}_{\rm J}$(n=1,2,3) states are predicted. The
radiative width to $4\,^3{\rm P}_{\rm J}$ multiplet is large, which
can provide access to the spin-triplet members of 4P multiplet.

\subsubsection{M1 transitions}

M1 transitions flip the quark spin, and M1 transitions are generally
suppressed relative to the E1 transitions, and it has been observed
in the charmonium system. M1 transition between different radial
multiplets are only nonzero due to the small relativistic
corrections to a vanishing lowest order M1 transition matrix
element, therefore there may be serious inaccuracy in some channels.
Analogous to the E1 transitions in the previous subsection, the M1
transitions width is given by\cite{Eichten:1974af,
Eichten:1975ag,Kwong:1988ae,Brambilla:2004wf}
\begin{equation}
\label{10} \Gamma_{{\rm M}1}({\rm n}^{2{\rm S}+1}{\rm L}_{\rm
J}\rightarrow {\rm n}'\,^{2{\rm S}'+1}{\rm L}'_{{\rm
J}'}+\gamma)=\frac{4\alpha e_c^2{\rm
E}^3_{\gamma}}{3m_c^2}\frac{2{\rm J}'+1}{2{\rm L}+1}\,\delta_{{\rm
L},{\rm L}'}\,\delta_{{\rm S},{\rm S}'\pm1}|\langle{\rm n}'\,^{2{\rm
S}'+1}{\rm L}'_{{\rm J}'}|j_0(\frac{{\rm E_{\gamma}}r}{2})|{\rm
n}\,^{2{\rm S}+1}{\rm L}_{\rm J}\rangle|^2\frac{{\rm E}_f}{{\rm
M}_i}
\end{equation}
where the meaning of the notations is the same as that in the E1
transition case. The above formula has included the recoil factor
$j_0({\rm E}_{\gamma}r/2)$ with $j_0(x)=\sin x/x$. Using the
wavefunctions of nonrelativistic potential model in Sec.II, the M1
transitions width both with and without the recoil factor are
calculated straightforwardly, theoretical predictions with the
corresponding photon energies are shown in Table
\ref{M1Y4361}--Table \ref{M1Y4704}. Obviously the M1 transitions of
Y(4360) and Y(4660) strongly depend on the recoil factors, and it
may be to small to be observed.

\section{strong decays of Y(4360) and Y(4660)}

Strong decays of mesons are driven by nonperturbative gluodynamics,
which are sensitive probe of hadron structure. However, it is very
difficult to be calculated from the first principle. For charmonium
above the ${\rm D}\overline{{\rm D}}$ threshold, the dominant decay
modes usually are the open charm strong decays, in which the initial
$c$ and $\bar{c}$ separate into different final states. OZI
forbidden decays are expected to be small, e.g. experimental
indications are that ${\cal B}(\psi(3770)\rightarrow
J/\psi\pi\pi)\sim 2.15\times10^{-3}-3.31\times10^{-3}$, hence we
shall focus on the open charm strong decays of Y(4360) and Y(4660)
in this section.

Although Open flavor decays are poorly understood from the QCD
dynamics so far, a number of phenomenological models have been
proposed to deal with this issue, the most popular are the $^3{\rm
P}_0$ model (quark pair creation
model)\cite{Micu:1968mk,orsay,Geiger:1994kr,Ackleh:1996yt}, the flux
tube model\cite{Isgur:1984bm,Kokoski:1985is} and the Cornell
model\cite{Eichten:1974af, Eichten:1975ag}. In the flux-tube model,
a meson consists of a quark and antiquark connected by a tube of
chromoelectric flux, which is treated as a vibrating string. For
conventional mesons the string is in its vibrational ground state.
The flux-tube breaking decay model\cite{Kokoski:1985is} is similar
to the $^3{\rm P}_0$ model, but extends it by including the dynamics
of the flux tubes. This is done by considering the overlap of the
flux tube of the initial meson with those of the two outgoing
mesons. $^3{\rm P}_0$ model is a limiting case of the flux tube
breaking model( $^3{\rm P}_0$ model emerges in the case of
infinitely thick flux tube ), which greatly simplifies the
calculations and gives similar results. The Cornell
model\cite{Eichten:1974af, Eichten:1975ag} assume that strong decays
take place through pair production from the linear confining
potential, which transform as the time component of a Lorentz vector
$j^{\,0}$, rather than the Lorentz scalar in the $^3{\rm P}_0$
model. The Cornell model has the advantage of unifying the
description of the spectrum and decays and completely specifies the
strength of the decay. Recently it has been used to discuss the
possible charmonium assignments of X(3872)\cite{Eichten:2004uh}.

The Orsay group have evaluated the open charm strong decays of three
$c\bar{c}$ states $\psi(3770)$, $\psi(4040)$ and $\psi(4415)$ in the
$^3{\rm P}_0$ model\cite{LeYaouanc:1977ux}, later these work was
extended by taking into account flux tube
breaking\cite{Page:1995rh}. Recently T.Barnes et al used the $^3{\rm
P}_0$ model to study the strong decays of both various candidates of
X(3872)\cite{Barnes:2003vb} and higher charmoinum up to the mass of
the 4S multiplet\cite{Barnes:2005pb}. In the following we shall
consider the open flavor strong decays of Y(4360) and Y(4660) as
$3\,^3{\rm D}_1$ and $5\,^3{\rm S}_1$ canonical charmonium in the
simple harmonic oscillator wavefunction approximation in the
framework of flux tube model, this approximation enables analytical
studies of amplitudes, and it is known to be an excellent
approximation for charmed mesons and light flavor mesons. Here we
assume the harmonic oscillator parameter $\beta$ of final states
mesons are identical, which is different from $\beta_A$ of the
initial charmoinum. We will calculate the decay width following the
procedure outlined in Ref.\cite{Kokoski:1985is,Ding:2006ya}.
Previous attempts on exploring the charmonium strong decay in
$^3{\rm P}_0$ model, flux tube model and Cornell model suggest that
the typical error of the partial width predictions is 30\%, and can
reach factors of 2 or even 3.

In the rest frame of $A$, the decay amplitude for an initial meson A
into two final mesons $B$ and $C$ is,
\begin{eqnarray}
\nonumber{\cal M}(A\rightarrow B+C)&&=\int d^{3}\mathbf{r}_A\int
d^{3}\mathbf{y}\;\psi_A(\mathbf{r}_A)\;\exp(i\frac{{\rm M}}{\rm m+M}\mathbf{p}_B\cdot\mathbf{r}_A)\;\gamma(\mathbf{r}_A,\mathbf{y})(i\nabla_{\mathbf{r}_B}+i\nabla_{\mathbf{r}_C}\\
\label{11}&&+\frac{2{\rm m}\;\mathbf{p_B}}{\rm
m+M})\,\psi^{*}_B(\mathbf{r}_B)\psi^{*}_C(\mathbf{r}_C)+(B\longleftrightarrow
C)
\end{eqnarray}
where both the flavor and spin overlap have been omitted in the
above amplitude, and $\gamma(\mathbf{r}_A,\mathbf{y})$ is the
flux-tube overlap function, which measures the spatial dependence of
the pair creation amplitude. $\mathbf{y}$ is the pair creation
position, $\mathbf{r}_A$, $\mathbf{r}_B$ and $\mathbf{r}_C$ are
respectively the quark-antiquark axes of $A$, $B$, and $C$ mesons,
they are related by $\mathbf{r}_B=\mathbf{r}_A/2+\mathbf{y}$,
$\mathbf{r}_C=\mathbf{r}_A/2-\mathbf{y}$. The initial
quark(antiquark) in $A$ is of mass ${\rm M}$ with ${\rm m}$ the mass
of the created quark pair. For charmonium decay concerned here,
${\rm M}=m_c$, ${\rm m}=m_{\rm q}$(q=u,d,s). When the flux tube is
in its ground states (conventional mesons), the flux-tube overlap
function is\cite{Kokoski:1985is}
\begin{equation}
\label{12}\gamma(\mathbf{r}_A,\mathbf{y})=A^{0}_{00}\;\sqrt{\frac{fb}{\pi}}\exp{(-\frac{fb}{2}\mathbf{y}^2_{\perp})}
\end{equation}

As usual, we take the string tension $b=0.18{\rm GeV}^2$, and the
constituent quark mass $m_u=m_d=0.33{\rm GeV}$, $m_s=0.55{\rm GeV}$
and $m_c=1.5{\rm GeV}$, and the estimated value $f=1.1$ and
$A^{0}_{00}=1.0$ are used in our calculation. The final ${\rm D}$
meson masses used to determined phase space and final state momentum
are taken from the Particle Data Group\cite{pdg} and from recent
Belle results\cite{Abe:2003zm}, and if not available, the estimated
mass motivated by the spectroscopy predictions are
used\cite{Godfrey:1985xj}. These masses are
$\rm{M}(\rm{D})=1.8694{\rm GeV}$, $\rm{M}(\rm{D}^{*})=2.0078{\rm
GeV}$, $\rm{M}(\rm{D}_0^{*})=2.308{\rm GeV}$(Belle),
$\rm{M}(\rm{D}_1)=2.444{\rm GeV}$, $\rm{M}(\rm{D}_1^{'})=2.422{\rm
GeV}$, $\rm{M}(\rm{D}_2)=2.459{\rm GeV}$,
$\rm{M}(\rm{D}^{'})=2.58{\rm GeV}$, $\rm{M}(\rm{D}^{*'})=2.64{\rm
GeV}$, $\rm{M}(\rm{D}_s)=1.9683{\rm GeV}$,
$\rm{M}(\rm{D}^{*}_s)=2.1121{\rm GeV}$,
$\rm{M}(\rm{D}^{*}_{s0})=2.317{\rm GeV}$,
$\rm{M}(\rm{D}_{s1})=2.459{\rm GeV}$,
$\rm{M}(\rm{D}^{'}_{s1})=2.535{\rm GeV}$,
$\rm{M}(\rm{D}_{s2})=2.572{\rm GeV}$, $\rm{M}(\rm{D}^{'}_s)=2.67{\rm
GeV}$, $\rm{M}(\rm{D}^{*'}_s)=2.73{\rm GeV}$.

Heavy-light mesons are not charge conjugation eigenstates and so
mixing can occur among states with the same $\rm{J}^{\rm P}$. The
$\rm{J}^{\rm P}=1^{+}$ axial vector $c\bar{n}$ and $c\bar{s}$ mesons
${\rm D}_1$ and ${\rm D}^{'}_1$ are the coherent superpositions of
quark model $^3{\rm P}_1$ and $^1{\rm P}_1$ states,
\begin{eqnarray}
\nonumber |{\rm D}_1\rangle&=&\cos\theta|^1{\rm
P}_1\rangle+\sin\theta|^3{\rm P}_1\rangle\\
\label{13}|{\rm D}^{'}_1\rangle&=&-\sin\theta|^1{\rm
P}_1\rangle+\cos\theta|^3{\rm P}_1\rangle
\end{eqnarray}

Little is known about the $^3{\rm P}_1$--\,$^1{\rm P}_1$ mixing
angle $\theta$ at present, however, in the heavy quark limit, the
mixing angle is predicted to be $-54.7^{\,o}$ or $35.3^{\,o}$ if the
expectation of heavy quark spin-orbit interaction is positive or
negative\cite{ Godfrey:1986wj,Godfrey:2005ww}. Since the former
implies that the $2^+$ state mass is larger than the $0^+$ state
mass, and this agrees with experiment, we assume
$\theta=-54.7^{\,o}$ in the following. We note that generally finite
quark mass will modify this mixing angle, and we can extract how
large the mixing angle is  by studying the dependence of the strong
decay amplitudes on the mixing angle $\theta$.

When we calculate the decay widths from the amplitudes, there are
ambiguities around the choice of phase space. The first choice is
the fully relativistic phase space(RPS), which leads to a factor of
$\frac{{\rm E}_{\rm B}{\rm E}_{\rm C}}{{\rm M}_{\rm A}}$ in the
final expression for the width in the center of mass frame, where
${\rm E_B}$ and ${\rm E_C}$ are respectively the energies of mesons
$\rm{B}$ and ${\rm C}$, and ${\rm M_A}$ is the mass of meson ${\rm
A}$. The second choice is fully non-relativistic phase space(NRPS),
then the energy factor is replaced by $\frac{{\rm M}_{\rm B}{\rm
M}_{\rm C}}{{\rm M}_{\rm A}}$, which is smaller than the
relativistic phase space. A third possibility employed by Kokoski
and Isgur, is the "mock meson" method, they suggest that the energy
factor should be $\frac{{\rm \widetilde{M}}_{\rm B}{\rm
\widetilde{M}}_{\rm C}}{{\rm \widetilde{M}}_{\rm A}}$, where ${\rm
\widetilde{M}}_i$(i=A,B,C) is the "mock meson" mass, which are the
calculated masses of the meson $i$ in the spin-independent
quark-antiquark potential\cite{Kokoski:1985is}. In practice, the
numerical result is little different from the relativistic phase
space except for the pseudoscalar goldstone bosons involved in the
final states. Therefore we shall give our partial width predictions
for the relativistic phase space(RPS) and non-relativistic phase
space(NRPS) in the following.

Theoretical estimates for the harmonic oscillator parameters $\beta$
and $\beta_{\rm A}$ scatter in a relative large region $0.3-0.7$
GeV. Many recent quark model studies of
mesons\cite{Ackleh:1996yt,Barnes:1996ff,Barnes:2002mu} and baryon
\cite{ Capstick:1993kb} decays in $^3{\rm P}_0$ model use the value
0.4GeV. Moreover, the harmonic oscillator parameters of ${\rm{D}}$,
${\rm D}^{*}$ and ${\rm D}(^3{\rm P}_{\rm J})$ etc are predicted to
be 0.45-0.66GeV, and mostly center around
0.5GeV\cite{Kokoski:1985is}. Therefore we take $\beta_{\rm
A}=0.4{\rm GeV}$, $\beta=0.5{\rm GeV}$ in our calculation as an
illustration, the outgoing mesons CM momentum $p_{\rm B}$, the
partial widths and strong decay amplitudes for the kinematically
allowed open charm decay modes of Y(4360) and Y(4660) are shown in
Table \ref{EY4360} --Table \ref{TY4660III}. We shall discuss some
interesting and characteristic aspects about the strong decays of
Y(4360) and Y(4660).

\subsection{Discussions about Y(4360) strong decay}

From Table \ref{EY4360}--Table \ref{TY4360II}, we can see that if
Y(4360) is a pure ${\rm 3\,^3D_1}$ $c\bar{c}$ state with mass 4.361
GeV, there are ten open charm strong decay modes: ${\rm DD, D^{*}D,
D^{*}D^{*}, D^{*}_2D, D^{*}_0D^{*}, D_1D, D^{'}D, D_sD_s,
D^{*}_sD_s, D^{*}_sD^{*}_s}$, and the total width is predicted to be
67.69 MeV(RPS) or 53.24 MeV(NRPS), comparing with the Belle
experimental measurement $74\pm15\pm10$ MeV. Provided that Y(4360)
mass is the prediction of non-relativistic potential model(4.455
GeV), then the additional decay modes ${\rm D_1D^{*}, D^{'}_1D^{*},
D_{s0}D^{*}_s\; and\; D_{s1}D_s}$ become available. There is
relative large difference between the relative phase space
normalization and the non-relativistic phase space normalization in
the $\rm DD$ mode, since the outgoing CM momentum $p_{\rm B}$ is
comparable to the D meson mass in this case.

The leading decay mode of Y(4360) is predicted to be $\rm DD$ with a
branching ratio$\approx57\%$, the second-largest decay mode is ${\rm
D^{*}D}$($\approx26\%$), and the ${\rm D^{*}D^{*}}$ mode also has
sizable branching ratio. The relative partial wave amplitudes in the
${\rm D^{*}D^{*}}$ final state are predicted to have a very
interesting pattern, ${\rm \frac{M_{P0}}{M_{P2}}=-\sqrt{5}}$, and
$\rm M_{F2}$ is predicted to be dominant, whereas it is zero for a
S-wave charmonium decay. Measuring the relative branching ratio
experimentally can determine whether Y(4360) is D-wave charmonium or
S-wave charmoium. In addition, we find the following relation,
\begin{eqnarray}
\nonumber  {\rm
M_{S1}(3\,^{3}D_{1}\rightarrow\,^{3}P_{1}+\,^{1}S_{0})}&=&
\frac{1}{\sqrt{2}}\,{\rm M_{S1}(3\,^{3}D_{1}\rightarrow\,^{1}P_{1}+\,^{1}S_{0})}\\
\label{14}{\rm M_{S1}(3\,^3D_1\rightarrow
\,^3P_1+\,^3S_1)}&=&\frac{1}{\sqrt{2}}\,{\rm
M_{S1}(3\,^3D_1\rightarrow \,^1P_1+\,^3S_1)}
\end{eqnarray}

Therefore for the heavy mixing angle $\theta=-54.7^o$, we have the
following relations,
\begin{equation}
\label{15}{\rm M_{S1}(Y(4360)\rightarrow
D_1+D)=M_{S1}(Y(4360)\rightarrow D_1+D^{*})}=0
\end{equation}

\begin{figure}[hptb]
\centering
\begin{minipage}[t]{0.46\textwidth}
\centering
\includegraphics[width=8cm]{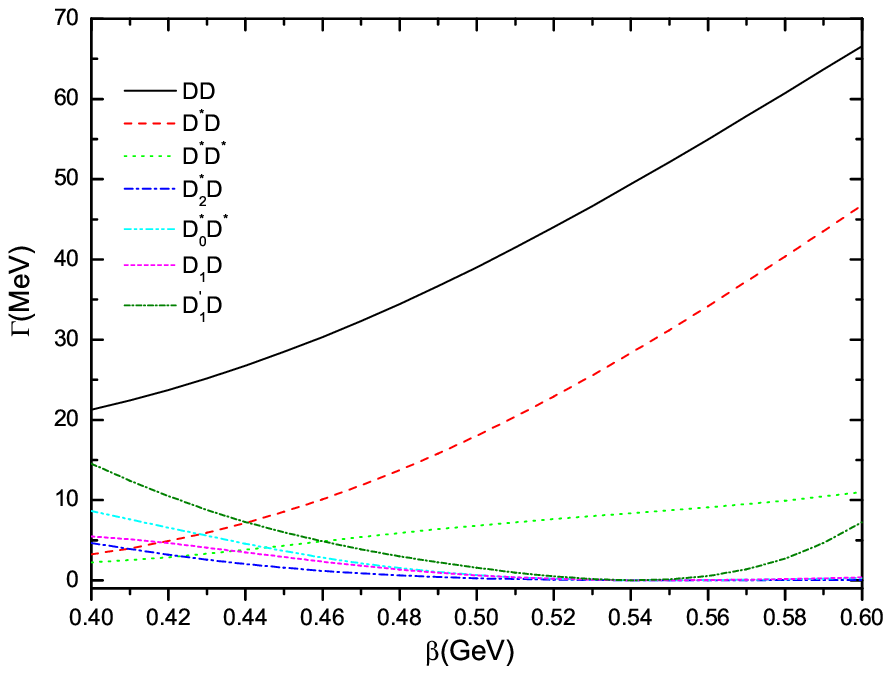}
\caption{\label{py4360}The variation of $\rm{DD, D^{*}D,
D^{*}D^{*},D^{*}_2D}$,$\rm{D^{*}_0D^{*},D_{1}D\;and\;D^{'}_1D}$
partial decay widths with $\beta$ for Y(4360) as a ${\rm 3\,^3{\rm
D_1} }$ charmonium state.}
\end{minipage}%
\hspace{0.04\textwidth}%
\begin{minipage}[t]{0.46\textwidth}
\centering
\includegraphics[width=8cm]{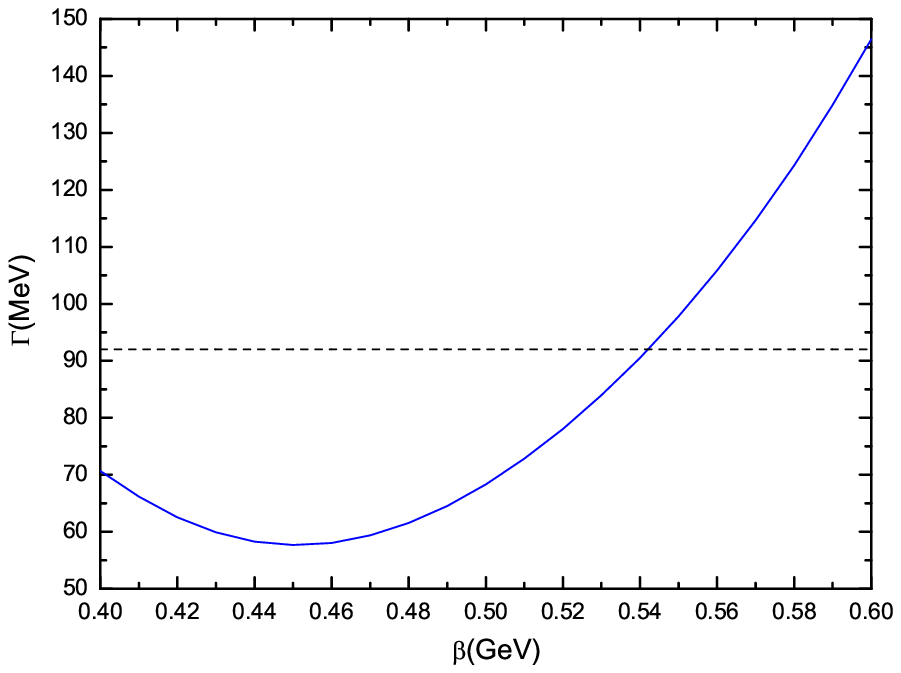}
\caption{\label{ay4360}Y(4360) total width dependence on $\beta$ as
a $3{\;^3{\rm D}_1}$ $c\bar{c}$ state in the flux tube model, the
horizontal line denotes the current experimental upper
bound\cite{Y4360}.}
\end{minipage}
\end{figure}

Thus, if Y(4360) is a pure $\rm 3\,^3D_1$ $c\bar{c}$ state, the
decays of Y(4360) to $\rm D_1D$ or $\rm D_1D^{*}$( if allowed by the
phase space ) are in D-wave rather than in S-wave, where $\rm D_1$
is the broader of the $1^+$ $c\bar{\rm{q}}$(q=u,d) axial vector
mesons. To test the robustness of our conclusions, we study the
stability of our results with respect to the variation of $\beta$.
The $\beta$ dependence of the partial decay width and the total
decay width are respectively shown in and Fig.\ref{py4360} and
Fig.\ref{ay4360}. In Fig.\ref{py4360} we showed the variation of
${\rm DD, D^{*}D,
D^{*}D^{*},D^{*}_2D,D^{*}_0D^{*},D_{1}D\;and\;D^{'}_1D}$ partial
decay widths with the harmonic oscillation parameter $\beta$, and we
see that the partial decay widths into S+P final states are small.

\subsection{Discussions about Y(4660) strong decay}

Since the mass of Y(4660) is large, many open charm strong decay
modes are allowable, which are listed obviously in Table
\ref{EY4660I}--Table \ref{TY4660III}. For Y(4660) mass being the
experimental value 4.664 GeV, the total width is predicted to be
45.04 MeV(RPS) or 32.78 MeV(NRPS) by our parameters, which is in
agreement with the Belle's measurement $48\pm15\pm3$ MeV. Y(4660)
dominantly decays into ${\rm D^{*}D,\,D^{*}D^{*}}$ with branching
ratios about $38\%$ and $34\%$ respectively, and $\rm DD$ is also an
important mode. Some partial width of S+P final states can be
comparable to the DD partial width for certain parameters. We would
like to mention that Y(4660) as a $5\,^3{\rm S_1}$ charmonium has
four nodes in the radial wavefunction, consequently some modes of
smaller branching ratios could be sensitive to the nodes positions.

It is interesting to note that the flux tube model predicts the
following relations between amplitudes,
\begin{eqnarray}
\nonumber{\rm
M_{S1}(5\,^3S_1\rightarrow\,^3P_1+\,^3S_1)}&=&-\sqrt{2}\,{\rm M_{S1}(5\,^3S_1\rightarrow\,^1P_1+\,^3S_1)}\\
\nonumber{\rm
M_{D1}(5\,^3S_1\rightarrow\,^3P_1+\,^3S_1)}&=&\frac{1}{\sqrt{2}}\,{\rm M_{D1}(5\,^3S_1\rightarrow\,^1P_1+\,^3S_1)}\\
\nonumber{\rm
M_{D2}(5\,^3S_1\rightarrow\,^3P_1+\,^3S_1)}&=&\frac{1}{\sqrt{2}}\,{\rm M_{D2}(5\,^3S_1\rightarrow\,^1P_1+\,^3S_1)}\\
\nonumber{\rm
M_{S1}(5\,^3S_1\rightarrow\,^3P_1+\,^1S_0)}&=&-\sqrt{2}\,{\rm M_{S1}(5\,^3S_1\rightarrow\,^1P_1+\,^1S_0)}\\
\label{16}{\rm
M_{D1}(5\,^3S_1\rightarrow\,^3P_1+\,^1S_0)}&=&\frac{1}{\sqrt{2}}\,{\rm
M_{D1}(5\,^3S_1\rightarrow\,^1P_1+\,^1S_0)}
\end{eqnarray}

then the following interesting relation appears,
\begin{eqnarray}
\nonumber {\rm M_{D1}(Y(4660)\rightarrow
D_1+D^{*})}&=&-\frac{1}{\sqrt{3}}\,{\rm M_{D2}(Y(4660)\rightarrow
D_1+D^{*})}\\
\label{17} {\rm M_{D1}(Y(4660)\rightarrow
D^{'}_1+D^{*})}&=&-\frac{1}{\sqrt{3}}\,{\rm
M_{D2}(Y(4660)\rightarrow D^{'}_1+D^{*})}
\end{eqnarray}

The above two relations are independent of the ${\rm ^3P_1-\,^1P_1}$
mixing angle $\theta$. For the heavy quark mixing angle
$\theta=-54.7^{\,o}$, we have the following relations,
\begin{eqnarray}
\nonumber &&{\rm M_{D1}(Y(4660)\rightarrow
D_1+D)=M_{D1}(Y(4660)\rightarrow
D_1+D^{*})=M_{D2}(Y(4660)\rightarrow D_1+D^{*})}=0\\
\label{18}&&{\rm M_{S1}(Y(4660)\rightarrow
D^{'}_1+D)=M_{S1}(Y(4660)\rightarrow D^{'}_1+D^{*})}=0
\end{eqnarray}

The above relations imply that Y(4660) decays into both ${\rm D_1D}$
and ${\rm D_1D^{*}}$ in S-wave, while into ${\rm D^{'}_1D}$ and
${\rm D{'}_1D^{*}}$ in D-wave, if it is purely a ${\rm 5\,^3S_1}$
$c\bar{s}$ state. These predictions can also be used to test whether
the ${\rm ^3P_1-\,^1P_1}$ mixing is consistent with the heavy quark
prediction.

As remarked in the previous discussion of Y(4360) decay, the ${\rm
D^{*}D^{*}}$ mode is especially interesting. There are four partial
wave amplitudes in this final state $\rm{M_{P0},M_{P1},
M_{P2},M_{F2}}$, both ${\rm M_{P1}}$ and ${\rm M_{F2}}$ are zero,
and the ratio of two nonzero P-wave amplitude is ${\rm
M_{P2}/M_{P0}=-2\sqrt{5}}$. However, this ratios is $-1/\sqrt{5}$ in
the case of D-wave charmonium decay, as is emphasized in the Y(4360)
decay. Experimentally measuring these ratios are essential for
understand the nature of Y(4660), and it is also a important test of
the flux tube decay model.

\begin{figure}[hptb]
\centering
\begin{minipage}[t]{0.46\textwidth}
\centering
\includegraphics[width=8cm]{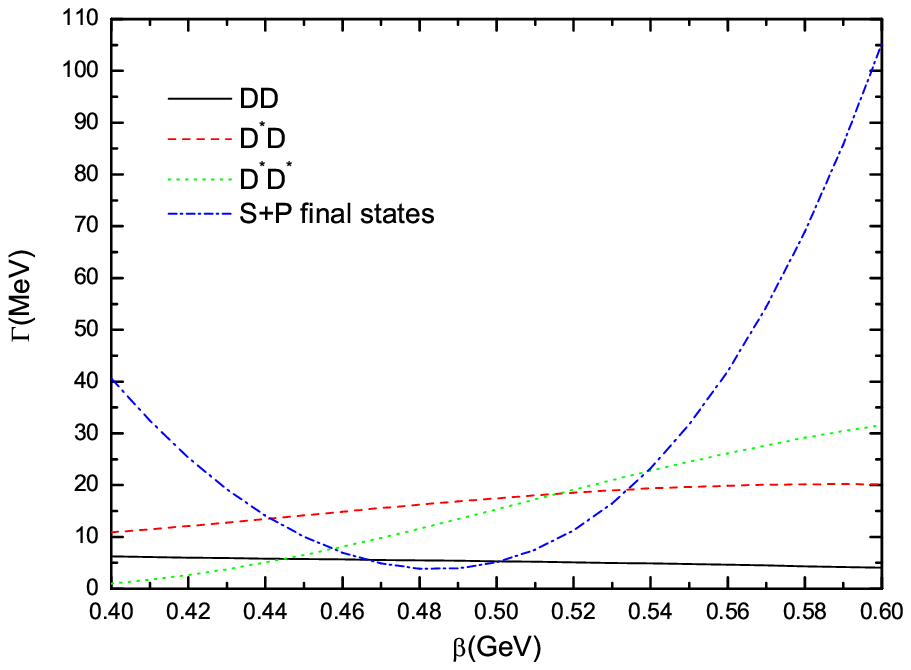}
\caption{\label{py4660}The variation of $\rm{DD, D^{*}D,
D^{*}D^{*}}$and total S+P final states partial decay widths with
$\beta$ for Y(4660) as a ${\rm 5\,^3{\rm S_1} }$ charmonium state.}
\end{minipage}%
\hspace{0.04\textwidth}%
\begin{minipage}[t]{0.46\textwidth}
\centering
\includegraphics[width=8cm]{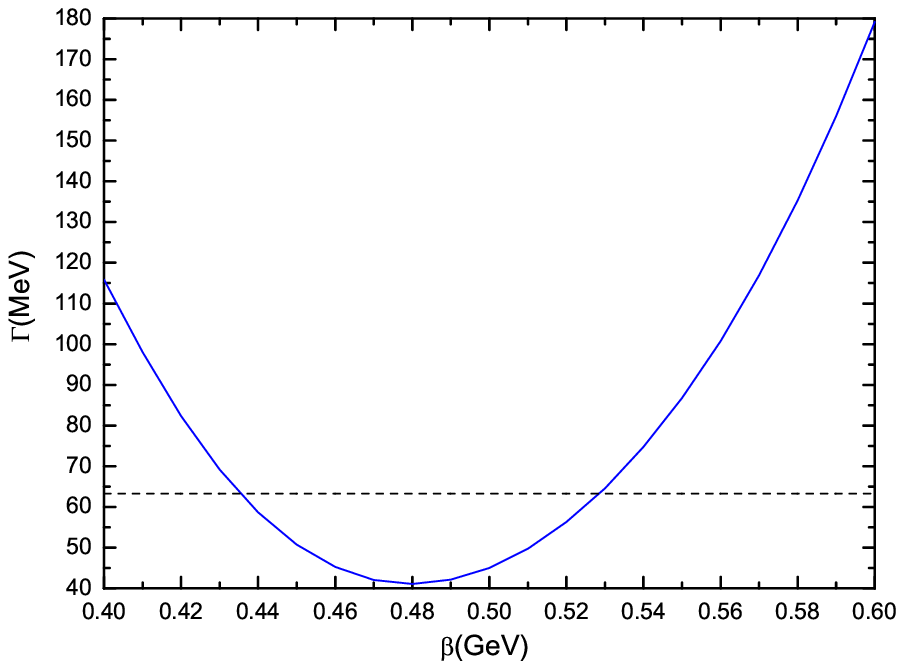}
\caption{\label{ay4660}Y(4660) total width dependence on $\beta$ as
a $5{\;^3{\rm S}_1}$ $c\bar{c}$ state in the flux tube model, the
horizontal line denotes the current experimental upper
bound\cite{Y4360}.}
\end{minipage}
\end{figure}

Moreover. Y(4660) can decay into ${\rm D^{*}_2D^{*}}$, which is
allowed by phase space. Five partial wave amplitudes are allowed for
this process ${\rm M_{S1},\rm M_{D1},\rm M_{D2},\rm M_{D3},\rm
M_{G3}}$, both ${\rm M_{S1}}$ and ${\rm M_{G3}}$ amplitudes are
predicted to be zero, whereas it is non-zero for a D-wave $c\bar{c}$
state decay. The ratios of the three D-wave amplitudes is ${\rm \rm
M_{D1}:\rm M_{D2}:\rm
M_{D3}}=1:\sqrt{\frac{5}{3}}:-4\sqrt{\frac{7}{3}}$. These
predictions can be used to test whether Y(4660) is a S-wave
charmonium, D-wave charmonium, or some other non-$c\bar{c}$
structure. In order to illustrate the parameter dependence of our
predictions, we show the $\beta$ dependence of the ${\rm DD, D^{*}D,
D^{*}D^{*}}$ and total S+P final states partial decay widths and the
total width in Fig.\ref{py4660} and Fig.\ref{ay4660} for Y(4660) as
a $5{\;^3\rm{S}_1}$ $c\bar{c}$ state. There are thirteen channels
whose final states are S-wave and P-wave D mesons, and each partial
decay width into S+P final state is at most close to the DD partial
width for large part of the parameters regions.

In short summary, if Y(4360) and Y(4660) are $3{\;^3{\rm D}_1}$ and
$5{\;^3{\rm S}_1}$ $c\bar{c}$ states respectively, the ${\rm DD,
D^{*}D\;and \;D^{*}D^{*}}$ are expected to be the dominant decay
modes, even if the variation of parameter is included. Some S+P
final state may has comparable branching ratio for certain
parameters. Careful measurements of these modes are crucial in
testing these charmonium assignments.

\section{discussion and conclusion}

In this paper, we consider the canonical charmonium assignments of
Y(4360) and Y(4660). Since these two structures are produced in
initial state radiation from $e^{+}e^{-}$ annihilation and hence to
have ${\rm J^{PC}}=1^{--}$, so they are S-wave or D-wave states if
they are canonical charmoniums. From the mass spectrum prediction of
the nonrelativistic potential model, we suggest Y(4360) is a ${\rm
3\,^3D_1}$ $c\bar{c}$ state, and $Y(4660)$ is a ${\rm 5\,^3S_1}$
$c\bar{c}$ state, if they are both conventional charmonium state,
although Y(4360) mass is somewhat smaller than the nonrelativistic
potential model prediction. We have investigated the $e^{+}e^{-}$
leptonic decay, E1 transions, M1 transitions and open charm strong
decay of both Y(4360) and Y(4660) in detail.

Although the mass of Y(4360) is consistent with Y(4325) observed by
the Babar collaboration, it is much narrower. Thus more data are
required to clarify whether they are the same structure. From the
$e^{+}e^{-}$ partial width of Y(4360), we estimate ${\cal
B}(Y(4360)\rightarrow\pi^{+}\pi^{-}\psi(2S))\sim1.20\times10^{-2}
(\;\rm{or}\;1.36\times10^{-2} )$, which is a little larger than the
corresponding branching ratio for a conventional D-wave $c\bar{c}$
state decay. It possibly may be due to large QCD radiative
corrections or some other non-$c\bar{c}$ componenets. It also
indicates we should examine other possible interpretations of
Y(4360) further, the ${\rm D_1\bar{D}}$ and ${\rm D_{s1}\bar{D}_s}$
threshold effects especially deserve considering
seriously\cite{progress}.

The Lattice QCD simulations predict that lightest charmonium
hybrid($c\bar{c}g$) is about
4.4GeV\cite{Bernard:1997ib,Mei:2002ip,Bali:2003tp}. It is obvious
that Y(4360) mass is very near to 4.4GeV, so we can not completely
exclude the possibility of Y(4360) as a $1^{--}$ charmonium hybrid,
although Y(4260) is already assumed to be a good candidate of
$1^{--}$ charmonium hybrid\cite{Close:2005iz,Zhu:2005hp,Kou:2005gt}.
Supposing that Y(4360) is a charmonium hybrid, its main decay modes
should be ${\rm D^*_2D, D^*_0D^*, D_1D,D^{'}_1D}$ according to the
famous "S+P" selection rule in hybrid decay, and the ${\rm
DD,D^*D,D^*D^*}$ modes should be highly suppressed. Consequently
measuring the ${\rm \rm DD,D^*D,D^*D^*, D^*_2D, D^*_0D^*,
D_1D,D^{'}_1D}$ modes are critical in distinguishing the canonical
charmonium from the charmonium hybrid interpretation.

$\chi_0(3\,^3{\rm P}_0)$, $\chi_1(3\,^3{\rm P}_1)$,
$\chi_0(2\,^3{\rm P}_0)$ and $\chi_1(2\,^3{\rm P}_1)$ are the main
Y(4360) E1 transition modes as a ${\rm 2\,^3D_1}$ $c\bar{c}$ state,
which possibly may be used to produce X(3940) and Y(3940), since
they are expected to belong ${\rm 2\,^3P_J}$ multiplet. The strong
suppression of Y(4660) E1 transitions to ${\rm n\,^{3}P_J}$(n=1,2,3)
multiplet is predicted. The M1 transition of Y(4360) and Y(4660)
should to too weak to be observed.

We have discussed the open charm strong decays of Y(4360) and
Y(4660) in the flux tube model in detail. Both Y(4360) and Y(4660)
are predicted to dominantly decay into ${\rm DD,\,D^{*}D,\,
D^{*}D^{*}}$, the partial width of some S+P final states can be
comparable with that of DD, ${\rm D^{*}D}$ or ${\rm D^{*}D^{*}}$ for
certain parameters. Measuring the ratios of the amplitudes in the
${\rm D^{*}D^{*}}$ final state will show whether Y(4360) and Y(4660)
are consistent with the charmonium assignments made in this work. If
Y(4360) is a pure $\rm 3\,^3D_1$ charmonium, ${\rm M_{F2}}$
amplitude is predicted to be largest and
$\rm{\frac{M_{P2}}{M_{P0}}}=-\frac{2}{\sqrt{5}}$. For Y(4660) as a
pure ${\rm 5\,^3S_1}$ $c\bar{c}$ state, we predict that the
amplitude $\rm M_{F2}$ is zero and
$\rm{\frac{M_{P2}}{M_{P0}}}=-2\sqrt{5}$. Similarly the ${\rm
D^*_2D^*}$ amplitude ratios in Y(4660) decay can test whether
Y(4660) is a S-wave or D-wave $c\bar{c}$ state, although the
branching ratio of ${\rm D^*_2D^*}$ is predicted to be small.
Provided that Y(4660) is a ${\rm 5\,^3S_1}$ $c\bar{c}$ state, we
have $\rm
M_{D1}:M_{D2}:M_{D3}=1:\sqrt{\frac{5}{3}}:-4\sqrt{\frac{7}{3}}$, and
the amplitudes $\rm M_{S1}$ and $\rm M_{G3}$ is zero, which is
non-zero for a D-wave state decay. The above results are generally
correct for S-wave or D-wave initial state decay. The careful
measurement of these relative branching ratios would play a critical
role in understanding Y(4360) and Y(4660).

The Belle and Babar Collaboration have measured the exclusive
$e^{+}e^{-}\rightarrow {\rm DD}$, $e^{+}e^{-}\rightarrow {\rm
DD^{*}}$ and $e^{+}e^{-}\rightarrow {\rm D^{*}D^{*}}$ cross section
using initial state
radiation\cite{Abe:2006fj,collaboration:2007mb,Collaboration:2007pa},
and the shapes of the cross sections are similar. There is a peak in
the Y(4660) region, however, no structure is clearly observed near
position of Y(4360) so far. Therefore Y(4660) as a $\rm 5\,^3S_1$
$c\bar{c}$ state is consistent with current experimental data,
however, Y(4360) may be a state beyond the quark model. Since the
${\rm DD_1}$ threshold is 4291.4 MeV (${\rm M_{D}+M_{D_1}\approx
4291.4 MeV}$), which is close to Y(4260), a possible way of
reconciling Y(4260) and Y(4360) is that Y(4260) is mainly the S-wave
${\rm DD_1}$ threshold effect and Y(4360) is a charmonium hybrid.
The relevant work is in progress\cite{ding}.

The confirmations and more experimental studies on Y(4360) and
Y(4660) at BES and CLEO are expected. Careful study of Y(4360) and
Y(4660) will greatly shed light on the charmonium spectroscopy.

\section*{ACKNOWLEDGEMENTS}
\indent  We acknowledge Prof. Dao-Neng Gao for very helpful and
stimulating discussions, and we are especially grateful to Prof.
Chang-Zheng Yuan for his professional comments. This work is
partially supported by National Natural Science Foundation of China
under Grant Numbers 90403021, 10005008, and KJCX2-SW-N10 of the
Chinese Academy.

\newpage

\begin{table}[hptb]
\caption{\label{E1Y4361}E1 radiative transitions of Y(4360)($3^3{\rm
D}_1$), and the Y(4360) mass is taken from experiment.}
\begin{ruledtabular}
\begin{tabular}{lcc}
Final meson & {E$_{\gamma}$ (MeV)} &{$\Gamma$ (keV)} \\\hline
$\chi_2(3\,^3{\rm P}_2)$ & 44 & 0.12\\
$\chi_1(3\,^3{\rm P}_1)$ & 89 & 15.9\\
$\chi_0(3\,^3{\rm P}_0)$ & 156 & 112\\ \\
$\chi_2(2\,^3{\rm P}_2)$ & 371 & 0.39\\
$\chi_1(2\,^3{\rm P}_1)$ & 414 & 8.00\\
$\chi_0(2\,^3{\rm P}_0)$ & 479 & 16.2\\
$\chi_2(2\,^3{\rm F}_2)$ & 10 & 0.062\\ \\
$\chi_2(1\,^3{\rm P}_2)$ & 731 & 0.17\\
$\chi_1(1\,^3{\rm P}_1)$ & 767 & 2.94\\
$\chi_0(1\,^3{\rm P}_0)$ & 843 & 5.11\\
$\chi_2(1\,^3{\rm F}_2)$ & 319 & 0.053
\end{tabular}
\end{ruledtabular}
\end{table}

\begin{table}
\caption{\label{E1Y4455}E1 radiative transitions of Y(4360)($3^3{\rm
D}_1$), and the Y(4360) mass is the prediction of the
nonrelativistic potential model, which is 4.455GeV.}
\begin{ruledtabular}
\begin{tabular}{lcc}
Final meson & {E$_{\gamma}$ (MeV)} &{$\Gamma$ (keV)} \\\hline
$\chi_2(3\,^3{\rm P}_2)$ & 135 & 3.65\\
$\chi_1(3\,^3{\rm P}_1)$ & 180 & 128\\
$\chi_0(3\,^3{\rm P}_0)$ & 246 & 425\\ \\
$\chi_2(2\,^3{\rm P}_2)$ & 456 & 0.71\\
$\chi_1(2\,^3{\rm P}_1)$ & 498 & 13.7\\
$\chi_0(2\,^3{\rm P}_0)$ & 562 & 25.6\\
$\chi_2(2\,^3{\rm F}_2)$ & 103 & 58.3\\ \\
$\chi_2(1\,^3{\rm P}_2)$ & 808 & 0.23\\
$\chi_1(1\,^3{\rm P}_1)$ & 844 & 3.85\\
$\chi_0(1\,^3{\rm P}_0)$ & 918 & 6.49\\
$\chi_2(1\,^3{\rm F}_2)$ & 405 & 0.11
\end{tabular}
\end{ruledtabular}
\end{table}

\begin{table}
\caption{\label{E1Y4664}E1 radiative transitions of Y(4660)($5^3{\rm
S}_1$), and the Y(4360) mass is taken from experiment.}
\begin{ruledtabular}
\begin{tabular}{lcc}
Final meson & {E$_{\gamma}$ (MeV)} &{$\Gamma$ (keV)} \\\hline
$\chi_2(4\,^3{\rm P}_2)$ & 42  & 7.92\\
$\chi_1(4\,^3{\rm P}_1)$ & 87 & 42.8\\
$\chi_0(4\,^3{\rm P}_0)$ & 152 & 75.5\\ \\
$\chi_2(3\,^3{\rm P}_2)$ & 334 & 0.34\\
$\chi_1(3\,^3{\rm P}_1)$ & 377 & 0.29\\
$\chi_0(3\,^3{\rm P}_0)$ & 439 & 0.15\\ \\
$\chi_2(2\,^3{\rm P}_2)$ & 640 & 0.68\\
$\chi_1(2\,^3{\rm P}_1)$ & 681 & 0.48\\
$\chi_0(2\,^3{\rm P}_0)$ & 741 & 0.20\\ \\
$\chi_2(1\,^3{\rm P}_2)$ & 976 & 0.37\\
$\chi_1(1\,^3{\rm P}_1)$ & 1010 & 0.24\\
$\chi_0(1\,^3{\rm P}_0)$ & 1082 & 0.097
\end{tabular}
\end{ruledtabular}
\end{table}

\begin{table}
\caption{\label{E1Y4704}E1 radiative transitions of Y(4660)($5^3{\rm
S}_1$), and the Y(4660) mass is the prediction of the
nonrelativistic potential model, which is 4.704GeV.}
\begin{ruledtabular}
\begin{tabular}{lcc}
Final meson & {E$_{\gamma}$ (MeV)} &{$\Gamma$ (keV)} \\
\hline
$\chi_2(4\,^3{\rm P}_2)$ & 81  & 57.7\\
$\chi_1(4\,^3{\rm P}_1)$ & 126 & 129\\
$\chi_0(4\,^3{\rm P}_0)$ & 191 & 147\\ \\
$\chi_2(3\,^3{\rm P}_2)$ & 371 & 0.46\\
$\chi_1(3\,^3{\rm P}_1)$ & 413 & 0.38\\
$\chi_0(3\,^3{\rm P}_0)$ & 475 & 0.19\\ \\
$\chi_2(2\,^3{\rm P}_2)$ & 675 & 0.78\\
$\chi_1(2\,^3{\rm P}_1)$ & 714 & 0.55\\
$\chi_0(2\,^3{\rm P}_0)$ & 774 & 0.23\\ \\
$\chi_2(1\,^3{\rm P}_2)$ & 1008 & 0.40\\
$\chi_1(1\,^3{\rm P}_1)$ & 1042 & 0.26\\
$\chi_0(1\,^3{\rm P}_0)$ & 1112 & 0.10
\end{tabular}
\end{ruledtabular}
\end{table}

\begin{table}[hptb]
\caption{\label{M1Y4361}M1 radiative transitions of
Y(4360)($3\,^3{\rm D}_1$), and Y(4360) mass is the experimental
value 4.361GeV.}
\begin{ruledtabular}
\begin{tabular}{lccc}
Final meson & {E$_{\gamma}$ (MeV)} & {$\Gamma$ (keV)} &
{$\Gamma_{rec}$ (keV)} \\\hline
$\eta_{c2}(2\,^1{\rm D}_2)$ & 199  & 0.00031 & 0.082\\
$\eta_{c2}(1\,^1{\rm D}_2)$ & 525  & 0.00067 & 0.19
\end{tabular}
\end{ruledtabular}
\end{table}

\begin{table}
\caption{\label{M1Y4455}M1 radiative transitions of
Y(4360)($3\,^3{\rm D}_1$), and Y(4360) mass is the theoretical
prediction of the nonrelativistic potential model, which is
4.455GeV.}
\begin{ruledtabular}
\begin{tabular}{lccc}
Final meson & {E$_{\gamma}$ (MeV)} & {$\Gamma$ (keV)} &
{$\Gamma_{rec}$ (keV)} \\\hline
$\eta_{c2}(2\,^1{\rm D}_2)$ & 287  & 0.00093 & 0.24\\
$\eta_{c2}(1\,^1{\rm D}_2)$ & 607  & 0.0010 & 0.29
\end{tabular}
\end{ruledtabular}
\end{table}

\begin{table}[hptb]
\caption{\label{M1Y4664}M1 radiative transitions of
Y(4660)($5\,^3{\rm{S}}_1$), and Y(4660) mass is the experimental
value 4.664GeV.}
\begin{ruledtabular}
\begin{tabular}{lccc}
Final meson & {E$_{\gamma}$ (MeV)} & {$\Gamma$ (keV)} &
{$\Gamma_{rec}$ (keV)} \\\hline
$\eta_c(4\,^1{\rm S}_0)$ & 272  & 0.15 & 0.95\\
$\eta_c(3\,^1{\rm S}_0)$ & 607  & 0.47 & 3.45\\
$\eta_c(2\,^1{\rm S}_0)$ & 913  & 0.82 & 4.26\\
$\eta_c(1\,^1{\rm S}_0)$ & 1381 & 2.65 & 9.36
\end{tabular}
\end{ruledtabular}
\end{table}

\begin{table}
\caption{\label{M1Y4704}M1 radiative transitions of
Y(4660)($5\,^3{\rm S}_1$), and Y(4660) mass is the theoretical
prediction of the nonrelativistic potential model, which is
4.704GeV} \vskip 0.3cm
\begin{ruledtabular}
\begin{tabular}{lccc}
Final meson & {E$_{\gamma}$ (MeV)} & {$\Gamma$ (keV)} &
{$\Gamma_{rec}$ (keV)} \\\hline
$\eta_c(5\,^1{\rm S}_0)$ & 19   & 0.013 & 0.013\\
$\eta_c(4\,^1{\rm S}_0)$ & 309  & 0.23 & 1.39\\
$\eta_c(3\,^1{\rm S}_0)$ & 642  & 0.55 & 4.05\\
$\eta_c(2\,^1{\rm S}_0)$ & 945  & 0.90 & 4.69\\
$\eta_c(1\,^1{\rm S}_0)$ & 1409 & 2.80 & 9.89
\end{tabular}
\end{ruledtabular}
\end{table}

\newpage

\begin{table}
\caption{\label{EY4360}Open-charm strong decay of
Y(4360)($3^3\rm{D}_1$), a factor of $+i$ has been suppressed in all
old partial waves. Y(4360) mass is the experimental value 4.361GeV.}
\begin{ruledtabular}
\begin{tabular}{lcccl}
Mode & {$p_{B}$ (GeV)} &\multicolumn{2}{c}{$\Gamma$ (MeV)} & {Amps(GeV$^{-1/2}$)}\\
     &                 &  RPS    &  NRPS &     \\\hline
$\rm{DD}$            &   1.12   &   39.03    &   28.69    & $\rm{M_{P0}}=0.3877$\\
$\rm{D^{*}D}$        &   1.00   &   18.02    &   14.24    & $\rm{M_{P1}}=-0.1977$\\
$\rm{D^{*}D^{*}}$    &   0.85   &   6.83     &   5.79     & $\rm{M_{P0}}=0.0705$\\
                     &          &            &            & $\rm{M_{P1}}=0$\\
                     &          &            &            & $\rm{M_{P2}}=-0.0315$\\
                     &          &            &            & $\rm{M_{F2}}=0.1695$\\
$\rm{D_2^{*}D}$      &   0.26   &   0.26     &   0.25     & $\rm{M_{D2}}=-0.0463$\\
$\rm{D_0^{*}D^{*}}$  &   0.31   &   0.67     &   0.66     & $\rm{M_{S1}}=0$\\
                     &          &            &            & $\rm{M_{D1}}=-0.0684$\\
$\rm{D_1D}$          &   0.32   &   0.62     &   0.60     & $\rm{M_{S1}}=0$\\
                     &          &            &            & $\rm{M_{D1}}=-0.0653$\\
$\rm{D^{'}_1D}$      &   0.38   &   1.56     &   1.50     & $\rm{M_{S1}}=0.0776$\\
                     &          &            &            & $\rm{M_{D1}}=-0.0534$\\
$\rm{D_sD_s}$        &   0.94   &   1.23     &   1.01     & $\rm{M_{P0}}=0.1066$\\
$\rm{D^{*}_sD_s}$    &   0.77   &   0.00     &   0.00     & $\rm{M_{P1}}=4.3\times10^{-4}$\\
$\rm{D_s^{*}D_s^{*}}$  & 0.54   &   0.11     &   0.10     & $\rm{M_{P0}}=-0.0135$\\
                       &        &            &            & $\rm{M_{P1}}=0$\\
                       &        &            &            & $\rm{M_{P2}}=0.0060$\\
                       &        &            &            & $\rm{M_{F2}}=-0.0384$\\
Total                  &        &   68.33    &    52.84   &
\end{tabular}
\end{ruledtabular}
\end{table}

\begin{table}
\caption{\label{TY4360I}Open-charm strong decay of
Y(4360)($3^3\rm{D}_1$), Y(4360) mass is 4.455GeV the prediction of
the nonrelativistic potential model.}
\begin{ruledtabular}
\begin{tabular}{lcccl}
Mode & {$p_{B}$ (GeV)} &\multicolumn{2}{c}{$\Gamma$ (MeV)} & {Amps(GeV$^{-1/2}$)}\\
     &                 &  RPS    &  NRPS &     \\\hline
$\rm{DD}$            &  1.21     &  51.33     &   36.15         & $\rm{M_{P0}}=0.4234$\\
$\rm{D^{*}D}$        &  1.10     &  35.42     &   26.81         & $\rm{M_{P1}}=-0.2615$\\
$\rm{D^{*}D^{*}}$    &  0.96     &  18.22     &   14.80         & $\rm{M_{P0}}=0.1411$\\
                     &           &            &                 & $\rm{M_{P1}}=0$\\
                     &           &            &                 & $\rm{M_{P2}}=-0.0631$\\
                     &           &            &                 & $\rm{M_{F2}}=0.2367$\\
$\rm{D_2^{*}D}$      &  0.52     &  0.02      &   0.02          & $\rm{M_{D2}}=-0.0089$\\
$\rm{D_0^{*}D^{*}}$  &  0.55     &  0.00      &   0.00          & $\rm{M_{S1}}=0$\\
                     &           &            &                 & $\rm{M_{D1}}=0.0034$\\
$\rm{D_1D^{*}}$      &  0.08     &  0.01      &   0.01          & $\rm{M_{S1}}=0$\\
                     &           &            &                 & $\rm{M_{D1}}=0.0117$\\
                     &           &            &                 & $\rm{M_{D2}}=0.0003$\\
$\rm{D_1D}$          &  0.55     &  0.00      &   0.00          & $\rm{M_{S1}}=0$\\
                     &           &            &                 & $\rm{M_{D1}}=0.0004$\\
$\rm{D^{'}_1D^{*}}$  &  0.24     &  0.50      &   0.49          & $\rm{M_{S1}}=-0.0323$\\
                     &           &            &                 & $\rm{M_{D1}}=-0.0304$\\
                     &           &            &                 & $\rm{M_{D2}}=-0.0506$\\
$\rm{D^{'}_1D}$      &  0.59     &  0.34      &   0.31          & $\rm{M_{S1}}=-0.0046$\\
                     &           &            &                 & $\rm{M_{D1}}=0.0345$\\
$\rm{D_sD_s}$        &  1.04     &  4.70      &   3.67          & $\rm{M_{P0}}=0.1953$\\
$\rm{D^{*}_sD_s}$    &  0.89     &  0.55      &   0.46          & $\rm{M_{P1}}=-0.0509$\\
$\rm{D_s^{*}D_s^{*}}$  & 0.71    &  0.02      &   0.02          & $\rm{M_{P0}}=-0.0117 $\\
                       &         &            &                 & $\rm{M_{P1}}=0$\\
                       &         &            &                 & $\rm{M_{P2}}=0.0052$\\
                       &         &            &                 & $\rm{M_{F2}}=0.0091$
\end{tabular}
\end{ruledtabular}
\end{table}

\begin{table}[hptb]
\caption{\label{TY4360II}Open-charm strong decay of
Y(4360)($3^3\rm{D}_1$), Y(4360) mass is 4.455GeV the prediction of
the nonrelativistic potential model(continued).}
\begin{ruledtabular}
\begin{tabular}{lcccl}
Mode & {$p_{B}$ (GeV)} &\multicolumn{2}{c}{$\Gamma$ (MeV)} & {Amps(GeV$^{-1/2}$)}\\
     &                 &  RPS    &  NRPS &     \\\hline
$\rm{D^{*}_{s0}D^{*}_s}$& 0.24   &  0.12      &   0.12          & $\rm{M_{S1}}=0$\\
                     &           &            &                 & $\rm{M_{D1}}=-0.0462$\\
$\rm{D_{s1}D_s}$     & 0.25      &  0.11      &   0.10          & $\rm{M_{S1}}=0$\\
                     &           &            &                 & $\rm{M_{D1}}=-0.0428$\\
Total                &           &  111.35    &   82.96         &
\end{tabular}
\end{ruledtabular}
\end{table}
\newpage

\begin{table}
\caption{\label{EY4660I}Open-charm strong decay of
Y(4660)($5^3\rm{S}_1$). ${\rm M}_{\rm LJ}$ is the partial wave
amplitude, where {\rm L=S, P, D,...} is the relative angular
momentum and J is their total spin. Note that a factor of $+i$ has
been suppressed in all old partial waves. Y(4660) mass is the
experimental value 4.664GeV.}
\begin{ruledtabular}
\begin{tabular}{lcccl}
Mode & {$p_{B}$ (GeV)} &\multicolumn{2}{c}{$\Gamma$ (MeV)} & {Amps(GeV$^{-1/2}$)}\\
     &                 &  RPS    &  NRPS &     \\\hline
$\rm{DD}$            &  1.39  &   5.29   &  3.40    & $\rm{M_{P0}}=0.1238$\\
$\rm{D^{*}D}$        &  1.30  &   17.47  &  12.06   & $\rm{M_{P1}}=0.1651$\\
$\rm{D^{*}D^{*}}$    &  1.19  &   15.32  &  11.36   & $\rm{M_{P0}}=0.0499$\\
                     &        &          &          & $\rm{M_{P1}}=0$\\
                     &        &          &          & $\rm{M_{P2}}=-0.2230 $\\
                     &        &          &          & $\rm{M_{F2}}=0$\\
$\rm{D_2^{*}D^{*}}$  &  0.67  &   0.23   &  0.21    & $\rm{M_{S1}}=0$\\
                     &        &          &          & $\rm{M_{D1}}=0.0042$\\
                     &        &          &          & $\rm{M_{D2}}=0.0055$\\
                     &        &          &          & $\rm{M_{D3}}=-0.0258$\\
                     &        &          &          & $\rm{M_{G3}}=0$\\
$\rm{D_2^{*}D}$      &  0.86  &   0.75   &  0.64    & $\rm{M_{D2}}=-0.0422$\\
$\rm{D_0^{*}D^{*}}$  &  0.88  &   1.00   &  0.86    & $\rm{M_{S1}}=-0.0480$\\
                     &        &          &          & $\rm{M_{D1}}=0$\\
$\rm{D_1D^{*}}$      &  0.69  &   0.22   &  0.20    & $\rm{M_{S1}}=-0.0256$\\
                     &        &          &          & $\rm{M_{D1}}=0$\\
                     &        &          &          & $\rm{M_{D2}}=0$\\
$\rm{D_1D}$          &  0.88  &   0.94   &  0.80    & $\rm{M_{S1}}=-0.0467$\\
                     &        &          &          & $\rm{M_{D1}}=0$\\
$\rm{D^{'}_1D^{*}}$  &  0.73  &   0.01   &  0.01    & $\rm{M_{S1}}=0$\\
                     &        &          &          & $\rm{M_{D1}}=-0.0032$\\
                     &        &          &          & $\rm{M_{D2}}=0.0056$\\
$\rm{D^{'}_1D}$      &  0.91  &   1.78   &  1.50    & $\rm{M_{S1}}=0$\\
                     &        &          &          & $\rm{M_{D1}}=0.0634$\\
$\rm{D^{*'}D^{*}}$   &  0.19  &   0.54   &  0.54    & $\rm{M_{P0}}=0.0166$\\
\end{tabular}
\end{ruledtabular}
\end{table}

\begin{table}
\caption{\label{EY4660II}Open-charm strong decay of
Y(4660)($5^3\rm{S}_1$). Y(4660) mass is the experimental value
4.664GeV(continued).}
\begin{ruledtabular}
\begin{tabular}{lcccl}
Mode & {$p_{B}$ (GeV)} &\multicolumn{2}{c}{$\Gamma$ (MeV)} & {Amps(GeV$^{-1/2}$)}\\
     &                 &  RPS    &  NRPS &\\\hline
$\rm{D^{*'}D^{*}}$     &        &          &          & $\rm{M_{P1}}=0$\\
                       &        &         &         & $\rm{M_{P2}}=-0.0743$\\
                       &        &         &         & $\rm{M_{F2}}=0$\\
$\rm{D^{*'}D}$         & 0.59   &  0.28   &  0.26   & $\rm{M_{P1}}=-0.0313$\\
$\rm{D^{'}D^{*}}$      & 0.42   &  0.06   &  0.05   & $\rm{M_{P1}}=0.0165$\\
$\rm{D^{'}D}$          & 0.69   &  0.04   &  0.04   & $\rm{M_{P0}}=0.0107$\\
$\rm{D_sD_s}$          & 1.25   &  0.50   &  0.35   & $\rm{M_{P0}}=0.0567$\\
$\rm{D^{*}_sD_s}$      & 1.13   &  0.39   &  0.30   & $\rm{M_{P1}}=0.0373$\\
$\rm{D_s^{*}D_s^{*}}$  & 0.99   &  0.00   &  0.00   & $\rm{M_{P0}}=0.0014$\\
                       &        &         &         & $\rm{M_{P1}}=0$\\
                       &        &         &         & $\rm{M_{P2}}=-0.0061$\\
                       &        &         &         & $\rm{M_{F2}}=0$\\
$\rm{D^{*}_{s2}}D_{s}$ &0.53    &  0.00   &  0.00   & $\rm{M_{D2}}=0.0031$\\
$\rm{D^{*}_{s0}}D^{*}_{s}$ &0.73&  0.06   &  0.05   & $\rm{M_{S1}}=0.0180$\\
                       &        &         &         & $\rm{M_{D1}}=0$\\
$\rm{D_{s1}}D^{*}_{s}$ &0.46    &  0.02   &  0.02   & $\rm{M_{S1}}=0.0121$\\
                       &        &         &         & $\rm{M_{D1}}=0$\\
                       &        &         &         & $\rm{M_{D2}}=0$\\
$\rm{D_{s1}}D_{s}$     &0.73    &  0.06   &  0.05   & $\rm{M_{S1}}=0.0180$\\
                       &        &         &         & $\rm{M_{D1}}=0$\\
$\rm{D^{'}_{s1}}D^{*}_{s}$ &0.20&  0.01   &  0.01   & $\rm{M_{S1}}=0$\\
                       &        &         &         & $\rm{M_{D1}}=0.0052$\\
                       &        &         &         & $\rm{M_{D2}}=-0.0090$\\
$\rm{D^{'}_{s1}}D_{s}$ & 0.60   &  0.02   &  0.02   & $\rm{M_{S1}}=0$\\
                       &        &         &         & $\rm{M_{D1}}=-0.0106$\\
$\rm{D^{'}_sD_s}$      & 0.24   &  0.05   &  0.05   & $\rm{M_{P0}}=0.0280$\\
Total                  &        &  45.04  &  32.78  &
\end{tabular}
\end{ruledtabular}
\end{table}

\begin{table}
\caption{\label{TY4660I}Open-charm strong decay of
Y(4660)($5^3\rm{S}_1$). ${\rm M}_{\rm LJ}$ is the partial wave
amplitude, where {\rm L=S, P, D,...} is the relative angular
momentum and J is their total spin. Note that a factor of $+i$ has
been suppressed in all old partial waves. Y(4660) mass is the
prediction of the nonrelativistic potential model, which is
4.704GeV.}
\begin{ruledtabular}
\begin{tabular}{lcccl}
Mode & {$p_{B}$ (GeV)} &\multicolumn{2}{c}{$\Gamma$ (MeV)} & {Amps(GeV$^{-1/2}$)}\\
 &                 &  RPS    &  NRPS &     \\\hline
$\rm{DD}$            & 1.43   & 5.13    &  3.24     & $\rm{M_{P0}}=0.1200$\\
$\rm{D^{*}D}$        & 1.33   & 19.82   &  13.45    & $\rm{M_{P1}}=0.1728$\\
$\rm{D^{*}D^{*}}$    & 1.23   & 20.93   &  15.25    & $\rm{M_{P0}}=0.0571$\\
                     &        &         &           & $\rm{M_{P1}}=0$\\
                     &        &         &           & $\rm{M_{P2}}=-0.2554$\\
                     &        &         &           & $\rm{M_{F2}}=0$\\
$\rm{D_2^{*}D^{*}}$  & 0.73   & 0.01    &  0.01     & $\rm{M_{S1}}=0$\\
                     &        &         &           & $\rm{M_{D1}}=0.0009$\\
                     &        &         &           & $\rm{M_{D2}}=0.0012$\\
                     &        &         &           & $\rm{M_{D3}}=-0.0057$\\
                     &        &         &           & $\rm{M_{G3}}=0$\\
$\rm{D_2^{*}D}$      & 0.91   & 2.00    &  1.68     & $\rm{M_{D2}}=-0.0666$\\
$\rm{D_0^{*}D^{*}}$  & 0.93   & 3.37    &  2.83     & $\rm{M_{S1}}=-0.0852$\\
                     &        &         &           & $\rm{M_{D1}}=0$\\
$\rm{D_1D^{*}}$      & 0.76   & 0.06    &  0.05     & $\rm{M_{S1}}=-0.0122$\\
                     &        &         &           & $\rm{M_{D1}}=0$\\
                     &        &         &           & $\rm{M_{D2}}=0$\\
$\rm{D_1D}$          & 0.93   & 3.19    &  2.67     & $\rm{M_{S1}}=-0.0834$\\
                     &        &         &           & $\rm{M_{D1}}=0$\\
$\rm{D^{'}_1D^{*}}$  & 0.79   & 0.12    &  0.11     & $\rm{M_{S1}}=0$\\
                     &        &         &           & $\rm{M_{D1}}=0.0089$\\
                     &        &         &           & $\rm{M_{D2}}=-0.0155$\\
$\rm{D^{'}_1D}$      & 0.96   & 3.54    &  2.93     & $\rm{M_{S1}}=0$\\
                     &        &         &           & $\rm{M_{D1}}=0.0865$
\end{tabular}
\end{ruledtabular}
\end{table}

\begin{table}
\caption{\label{TY4660II}Open-charm strong decay of
Y(4660)($5^3\rm{S}_1$), Y(4660) mass is 4.704GeV the prediction of
the nonrelativistic potential model(continued).}
\begin{ruledtabular}
\begin{tabular}{lcccl}
Mode & {$p_{B}$ (GeV)} &\multicolumn{2}{c}{$\Gamma$ (MeV)} & {Amps(GeV$^{-1/2}$)}\\
  &                 &  RPS    &  NRPS &\\\hline
$\rm{D^{*'}D^{*}}$     & 0.36    & 0.01   &  0.01   & $\rm{M_{P0}}=0.0013$\\
                       &         &        &         & $\rm{M_{P1}}=0$\\
                       &         &        &         & $\rm{M_{P2}}=-0.0058$\\
                       &         &        &         & $\rm{M_{F2}}=0$\\
$\rm{D^{*'}D}$         & 0.66    &  0.00  &  0.00   & $\rm{M_{P1}}=-0.0029$\\
$\rm{D^{'}D^{*}}$      & 0.52    &  0.35  &  0.33   & $\rm{M_{P1}}=0.0372 $\\
$\rm{D^{'}D}$          & 0.75    &  0.69  &  0.62   & $\rm{M_{P0}}=0.0433$\\
$\rm{D_sD_s}$          & 1.29    &  0.70  &  0.49   & $\rm{M_{P0}}=0.0659$\\
$\rm{D^{*}_sD_s}$      & 1.17    &  0.76  &  0.57   & $\rm{M_{P1}}=0.0509$\\
$\rm{D_s^{*}D_s^{*}}$  & 1.03    &  0.06  &  0.05   & $\rm{M_{P0}}=0.0048$\\
                       &         &        &         & $\rm{M_{P1}}=0$\\
                       &         &        &         & $\rm{M_{P2}}=-0.0217$\\
                       &         &        &         & $\rm{M_{F2}}=0$\\
$\rm{D^{*}_{s2}}D^{*}_{s}$ & 0.22&  0.01  &  0.01   & $\rm{M_{S1}}=0$\\
                       &         &        &         & $\rm{M_{D1}}=-0.0026$\\
                       &         &        &         & $\rm{M_{D2}}=-0.0034$\\
                       &         &        &         & $\rm{M_{D3}}=0.0160$\\
                       &         &        &         & $\rm{M_{G3}}=0$\\
$\rm{D^{*}_{s2}}D_{s}$ & 0.61    &  0.02  &  0.02   & $\rm{M_{D2}}=0.0113$\\
$\rm{D^{*}_{s0}}D^{*}_{s}$ & 0.79 & 0.05  &  0.04   & $\rm{M_{S1}}=0.0159$\\
                       &         &        &         & $\rm{M_{D1}}=0$\\
$\rm{D_{s1}}D^{*}_{s}$ &  0.55   & 0.00   &  0.00   & $\rm{M_{S1}}=-4.3\times10^{-5}$\\
                       &         &        &         & $\rm{M_{D1}}=0$\\
                       &         &        &         & $\rm{M_{D2}}=0$\\
$\rm{D_{s1}}D_{s}$     &  0.79   & 0.05   &  0.04   & $\rm{M_{S1}}=0.0160$\\
                       &         &        &         & $\rm{M_{D1}}=0$
\end{tabular}
\end{ruledtabular}
\end{table}

\begin{table}
\caption{\label{TY4660III}Open-charm strong decay of
Y(4660)($5^3\rm{S}_1$), Y(4660) mass is 4.704GeV the prediction of
the nonrelativistic potential model(continued).}
\begin{ruledtabular}
\begin{tabular}{lcccl}
Mode & {$p_{B}$ (GeV)} &\multicolumn{2}{c}{$\Gamma$ (MeV)} & {Amps(GeV$^{-1/2}$)}\\
  &                 &  RPS    &  NRPS &\\\hline
$\rm{D^{'}_{s1}}D^{*}_{s}$& 0.36 & 0.02   &  0.02   & $\rm{M_{S1}}=0$\\
                       &         &        &         & $\rm{M_{D1}}=0.0073$\\
                       &         &        &         & $\rm{M_{D2}}=-0.0126$\\
$\rm{D^{'}_{s1}}D_{s}$ & 0.67    & 0.04   &  0.03   & $\rm{M_{S1}}=0$\\
                       &         &        &         & $\rm{M_{D1}}=-0.0149$\\
$\rm{D^{*'}_sD_s}$     & 0.11    & 0.03   &  0.03   & $\rm{M_{P1}}=0.0314$\\
$\rm{D^{'}_sD_s}$      & 0.39    & 0.00   &  0.00   & $\rm{M_{P0}}=0.0051$\\
Total                  &         & 60.96  &  44.48  &
\end{tabular}
\end{ruledtabular}
\end{table}

\end{document}